\documentclass[fleqn,usenatbib]{mnras}
\usepackage{newtxtext,newtxmath}
\usepackage[T1]{fontenc}
\usepackage{xcolor}
\DeclareRobustCommand{\VAN}[3]{#2}
\let\VANthebibliography\thebibliography
\def\thebibliography{\DeclareRobustCommand{\VAN}[3]{##3}\VANthebibliography}

\usepackage{graphicx}	
\usepackage{amsmath}	

\title[Constraining the parameters of SMNS via GW and EM]{Constraining the  ellipticity and frequency of  binary neutron star remnant via its gravitational-wave and electromagnetic radiations}

\author[Yong Yuan et al.]{
Yong Yuan,$^{1}$
Xi-Long Fan,$^{1}$\thanks{E-mail: xilong.fan@whu.edu.cn}
and Hou-Jun L\"{u}$^{2}$
\\
$^{1}$School of Physics Science And Technology, Wuhan University, No.299 Bayi Road0, Wuhan, Hubei, China\\
$^{2}$Guangxi Key Laboratory for Relativistic Astrophysics, School of Physical Science and Technology, Guangxi University, Nanning, Guangxi, China
}

\date{Accepted XXX. Received YYY; in original form ZZZ}

\pubyear{2022}

\begin{document}
\label{firstpage}
\pagerange{\pageref{firstpage}--\pageref{lastpage}}
\maketitle

\begin{abstract}
The nature of the merger remnant of binary neutron star (BNS) remains an open question. From the theoretical point of view, one possible outcome is a supra-massive neutron star (SMNS), which is supported by rigid rotation and through its survival of hundreds of seconds before collapsing into a black hole (BH). If this is the case, the SMNS can emit continuous gravitational waves (GW) and electromagnetic (EM) radiation, particularly in the X-ray band. In this work,  the ellipticity and initial frequency of SMNS are constrained  with a Bayesian framework using simulated X-ray and GW signals, which could be detected by The Transient High Energy Sky and Early Universe Surveyor (THESEUS) and Einstein Telescope (ET), respectively.
We found that only considering the X-ray emission can not completely constrain the initial frequency and ellipticity of the SMNS, but it can reduce the ranges of the parameters. Afterwards, we can use the posterior distribution of the X-ray parameter estimates as a prior for the GW parameter estimates. It was found that the 95$\%$ credible region of the joint X-ray–GW  analysis was about $10^5$ times smaller than that of the X-ray analysis alone.
\end{abstract}




\section{Introduction}\label{sec:intro}

On 2017 August at 17 12:41:04.4 UTC, the first observation of a BNS coalescence by the Advanced LIGO and Advanced Virgo gravitational-wave detectors \citep{AbbottPRL}, accompanied by the electromagnetic signals observed by the Fermi Gamma-ray Burst Monitor \citep{Goldstein2017ApJL, Savchenko2017ApJL}, ushered in the age of multi-messenger astronomy. Unfortunately, the LVC limited the post-merger GW amplitude combined with the X-ray signal, and there is still debate about what the remnant of a BNS merger might be \citep{2017ApJLAbbott, Goldstein2017ApJL, Abbott2019ApJ}. From a theoretical point of view, there are four different scenarios for BNS merger: (i) Form a black hole (BH) \citep{1999ApJPopham, 2000ApJWheeler, 2009ApJLei, 2017NewARLiu}. The GW quasi-normal mode ringdown signal from a remnant BH has a dominant frequency of around 6 kHz (the BNS mass is similar to GW170817) \citep{Shibata2006PhRvD, Baiotti2008PhRvD}, (ii) Form a hyper-massive neutron star (HMNS). It has a mass greater than the maximum mass of a uniformly rotating star. An HMNS is unlikely to survive for longer than a few tens to hundreds of milliseconds after a merger, before collapsing to a BH \citep{Shapiro2000ApJ, Hotokezaka2013PhRvD}, whereas a surviving HMNS generates strong peaks in the GW spectrum in the frequency range of 2-4 kHz due to sustained global oscillations \citep{1994PhRvDXing, 1996A&ARuffert,2000PhRvDShibata}, (iii) Form a SMNS. If the equation of state (EoS) of NS is stiff enough after the early violent evolution of the post-merger, the frequency of the SMNS dropped below 1 kHz \citep{2000A&ARosswog, 2006SciDai, 2006MNRASFan, 2013ApJLYu, Lv2015ApJ, 2016PhRvDGao}, (iv) Or even form a stable NS \citep{Dai1998a, Dai1998PhRvL}. In scenario (iii) and (iv) , the GW signals belong to the category of continuous gravitational wave signals. In scenario (iv), the frequency of the GW is closer to the monochromatic frequency, and it lasts longer than in scenario (iii). 

SMNS can not only produce gravitational waves, but also can emit electromagnetic waves in various wavelength bands \citep{Dai1998a, Zhang2001ApJ, Corsi2009ApJ, 2006SciPrice, Giacomazzo2013ApJL, Metzger2014MNRAS, 2016ARNPSRodrigo, Sowell2019Prd, Lv2020ApJL}. In terms of X-ray observations, the X-ray internal plateau (a fairly constant emission followed by a steep decay with a decay slope $t^{-\alpha}$, with $\alpha > 3$) in some short GRBs may be as a "smoking gun" for a supra-massive magnetar as the central engine \citep{Rowlinson2010MNRAS, Rowlinson2013MNRAS, Lv2015ApJ, Lv2017ApJ}. If the plateau emission followed by a $\sim t^{-1}$ decay phase, then followed by a deep decay phase $t^{-\alpha}$ with $\alpha >3$, the phenomenon implies that the SMNS mainly losses rotational energy through gravitational wave radiation, and then collapses into a BH \citep{Zhang2001ApJ, 2006SciDai, Gao2006ChJAA, Zhang2013ApJL, 2018MNRASLv, Sarin2020Prd, Sarin2020MNRAS}, such as GRB 050724, GRB 160821B and GRB 200219A \citep{Barthelmy2005Natur, Lv2017ApJ, Lv2020ApJL}. On the other hand, some researchers also claimed that the internal plateau emission in X-ray is not only interpreted as SMNS collapse into black hole, such as structured jets at large viewed angles  \citep{Beniamini2020MNRAS, Oganesyan2020ApJ}. In this paper, we assume that the internal plateau is from the SMNS collapse into black hole to perform the joint X-ray–GW data analysis.

According to the current sensitivity of Advanced LIGO and Advanced Virgo, the post-merger signal of BNS is very weak relative to the noise and its frequency beyond the LIGO sensitivity range \citep{AbbottPRL, 2017ApJLAbbott}, so it is difficult to search for the post-merger signal to determine which of the above-mentioned products is the final remnant. In the near future, with the improved sensitivity of the third generation of gravitational-wave detectors, such as Einstein Telescope (ET) \citep{2008Hild_ET} and Cosmic Explorer (CE) \citep{2017CQGraAbbott_CE}, there is a great chance to detect post-merger GWs generated by the merger of BNS, then we will be able to directly determine what kind of remnant is after the merger \citep{Brady2000PhRvD, Cutler2005PhRvD, Pletsch2009PhRvL,Pitkin2011MNRAS, Pletsch2010PhRvD, Pletsch2011PhRvD}.


Over the past decade, multi-messenger astrophysics as a distinct discipline providing unique and valuable insights into the properties and processes of the physical universe. Here, we only briefly introduce some works on joint GW and EM, more details can be found in e.g. \cite{Fan2015arXiv, Meszaros2019NatRP, Engel2022arXiv}. By combining the GW and EM information, the radius and maximum mass of NS can be constrained, thus the equation of state (EOS) of NS can be well constrained \citep{Margalit2017ApJL, Dietrich2020Sci}. The joint observations of GW and host galaxies can improvements to the source sky location estimates, and a better estimate of the inclination angle of the GW source can be obtained \citep{Fan2014ApJ, Chatterjee2023}. Combined GW and GRB can break the degeneration of distance and inclination angle \citep{Fan2017PhRvL, Guidorzi2017ApJ}, can constrain and distinguish the jet model of GRB \citep{Hayes2020ApJ}, and can also constrain the Hubble constant more precisely \citep{Guidorzi2017ApJ}. The joint detections of GW and kilonova can reveal the origin of heavy elements and also put a limit on the Hubble constant \citep{Metzger2017LRR,Kasen2017Natur,Kawaguchi2018ApJL}. The joint observations of GW and X-ray can effectively reduce the prior range of GW parameters, so as to accurately target search for GW \citep{Corsi2009ApJ, Sarin2018PhRvD}. The search algorithms for intermediate-duration GW  by  combing electromagnetic waves are also proposed in \cite{Coyne2016Prd, Sowell2019Prd}.


In this paper, we assume that SMNS forms after the merger of binary neutron stars (scenario (iii) discussed above), and the distance of the source can be measured in other ways. We simulate GW data with ET and X-ray data with THESEUS \citep{Amati2018}, to estimate the initial frequency and ellipticity of the NS. For an isolated neutron star, knowing its initial frequency and ellipticity helps us have a more direct and clear understanding of the neutron star's tidal deformation and  EoS \citep{FlanaganPRD2008, 1998ApJAndersson, 1998PhRvLLindblom} and we can constrain the magnetic field inside the star and oscillation mode \citep{Haskell2008MNRAS, Lasky2015PASA}. Although the initial frequency and ellipticity are coupled together in the parameter estimation of X-ray, it can narrow the ranges of parameters \citep{Corsi2009ApJ, Sarin2018PhRvD}, save computational resources, and improve the accuracy for the parameter estimation of GW. Finally, we simulate 100 samples to compare the 95$\%$ credible intervals on the ellipticity and initial frequency posterior distribution obtained using X-ray observations alone with joint X-ray–GW  observations.

The paper is set out as follows: in Section {\ref{sec:method}}, We introduce the theoretical model and statistical framework, in Section {\ref{sec:gw}}, we introduce the GW radiation model, and in Section \ref{sec:em}, the X-ray model will be described. In Section {\ref{sec:result}}, we simulate observation data and perform the joint parameter estimates for the initial frequency and ellipticity of the neutron star. Conclusions are drawn in Section {\ref{sec:summary}} with some additional discussion.

\section{Model} \label{sec:method}

\subsection{Gravitational Wave Waveforms}
\label{sec:gw}


SMNS spin-dwon is determined by the combination of magnetic dipole and GW quadrupole losses \citep{1983JBAAShapiro, Zhang2001ApJ, Fan2013PhRvD, Lasky2016MN, Abbott2019ApJ}:

\begin{equation}
    \frac{dE}{dt} = -\frac{B^2R^6\Omega^4}{6c^3} -
    \frac{32GI^2\epsilon^2\Omega^6}{5c^5} = L_{EM} + L_{GW}
    \label{Dedt}
\end{equation}
where $B$ is the star's dipolar field strength at the poles, $R$ is the mean stellar radius, $\Omega$ is the star's angular frequency, $\epsilon$ is the ellipticity, and $I$ is the moment of inertia with respect to the rotation axis. In general, the spin-down of SMNS can be well-described by the general torque equation:
\begin{equation}
    \dot{\Omega} = -k\Omega^n,
    \label{eq:torque}
\end{equation}
where an overdot represents a derivative with respect to time, $k$ is a constant of proportionality, and $n$ is the braking index. If the rotational energy loss is dominated by GW radiation, then $n=5$ \citep{2017ApJLAbbott, Lasky2017ApJ, Lv2019ApJ}. In this case, the first term on the right-hand side in Eq (\ref{Dedt}) is negligible, then
\begin{equation}
    \Omega = \Omega_0\left(1+\frac{128GI\epsilon^2\Omega_0^4}{5c^5}t\right)^{-1/4}
    \label{eq:Omega}
\end{equation}
where $\Omega_0$ is the initial angular frequency of the star, $\Omega_0 = 2\pi f_0$, $f_0$ is the initial frequency of the star, and GW radiation at a frequency equal to twice the rotation frequency. The GW strain can be given \citep{1979PhRvDZimmermann},

\begin{equation}
 h_+(t) = \left\{
\begin{array}{rcl}
\frac{4GI\epsilon \Omega^2}{c^4 d}\frac{1+ \cos^2{\iota}}{2} \cos{2 \Omega t} & & {t\leq t_{col}}\\
0 & & {t>t_{col}}\\
\end{array}\right.
\label{eq:hp}
\end{equation}

\begin{equation}
h_\times(t) = \left\{
\begin{array}{rcl}
\frac{4GI\epsilon \Omega^2}{c^4 d} \cos{\iota} \sin{2\Omega t} & & {t\leq t_{col}}\\
0 & & {t>t_{col}}\\
\end{array}\right.
\label{eq:hc}
\end{equation}
where $\iota$ is the inclination angle defined as the angle between the direction of the view and the rotation axis. $d$ is the luminosity distance and $t_{col}$ is the time scale at which the SMNS collapses into a BH. 

\subsection{X-ray internal plateau}
\label{sec:em}

SMNS as the central engine of the short Gamma Ray Burst (sGRB) has been discussed by \cite{Rowlinson2010MNRAS, Rowlinson2013MNRAS, Lv2015ApJ, 2018MNRASLv, Lv2019ApJ, Lv2020ApJL}.  
\cite{Lasky2016MN} and \cite{2018MNRASLv}  have derived the time-dependent EM luminosity for energy loss dominated by both EM dipole and GW emission. The resulting evolution patterns exhibit the following behaviors,
\begin{equation}
L_{EM}(t) = \left\{
\begin{array}{rcl}
L_{em,0}\left(1 + \frac{t}{\tau}\right)^{\frac{4}{1-n}} & & {t\leq t_{col}}\\
L_{em,0}\left(1 + \frac{t_{col}}{\tau}\right)^{\frac{4}{1-n}}\left(\frac{t}{t_{col}}\right)^{-\alpha} & & {t>t_{col}}\\
\end{array}\right.
\label{eq:Lem}
\end{equation}
where $L_{em,0}$ represents the initial luminosity of EM and $\alpha$ is the decay slopes after $t_{col}$. When the EM radiation is dominant, n=3, $\tau = \frac{3c^3I}{B^2R^6\Omega_0^2}$, while GW radiation is dominant, n=5,  $\tau = \frac{5c^5}{128GI\epsilon^2\Omega_0^4}$.

In terms of observation,  there is a plateau in X-ray evolution for many sGRBs (e.g. \cite{Barthelmy2005Natur, Lv2017ApJ, Lv2020ApJL}) following $L_{\text{X-ray}} \propto t^{-1}$, then a deep decay phase $L_{\text{X-ray}} \propto t^{-\alpha}$ $(\alpha > 3)$. This type of  X-ray light curve evolution (e.g. the case shown in  lower panel of Fig \ref{fig:data}) is consistent with Eq. (\ref{eq:Lem}) and  the scenario (iii) discussed above,  and can be understood as the typical emission  from an SMNS dominated by GW radiation, losing rotation energy for about a few hundred seconds, and then collapsing to BH.  In this case, it implies that $n=5$. The X-ray flux can be expressed as:

\begin{equation}
    F(t) = \eta\frac{L_{EM}(t)}{4\pi d^2},
    \label{flux}
\end{equation}
where $\eta = \frac{L_{\text{X-ray}}}{L_{EM}}$ \citep{Lv2020ApJL}. Additionally, it is believed that after $\Delta t$, the SMNS will collapse into a BH at $\tau$,
\begin{equation}
t_{col} = \tau + \Delta t.
\end{equation}

\subsection{Data analysis method}

We aim to obtain posterior distributions of ellipticity and initial by joint X-ray–GW  observations. According to Eq (\ref{eq:hp}) and (\ref{eq:hc}), the GW simulation observation $\mathcal{S}$ can be generated by Eq (\ref{eq:st}),
\begin{equation}
    \mathcal{S} = F_+ h_+(t) + F_\times h_\times(t) + n_g(t)
    \label{eq:st}
\end{equation}
where $F_+$ and $F_\times$ represent the antenna pattern functions and $n_g(t)$ is the instrument noise. Based on Eq (\ref{flux}), the X-ray simulation observation $\mathcal{H}$ can be generated by Eq (\ref{eq:Ht}),
\begin{equation}
\mathcal{H} = F(t) + n_x(t)
\label{eq:Ht}
\end{equation}
where the $n_x(t)$ is the noise of the X-ray signal.

By Eq. (\ref{eq:st}) and Eq. (\ref{eq:Ht}), we can get the observation sets of GW and X-ray, and the observation parameters of GW and X-ray are divided into three sets. The set of common parameters of observations is denoted by $\gamma = (\epsilon, f_0)$. The parameters that are distinct to either only the X-ray or GW observations are denoted by $\theta = (\eta, \alpha, n, \Delta t, L_{em,0})$ and $\phi=(I, \iota)$ respectively. Any other implicit model assumptions denoted by $M$. We use Bayes' theorem for parameter estimation while ignoring the evidence, the posterior distribution of X-ray $p_x$ and GW $p_g$ about $\gamma$ can be written as
\begin{equation}
p_x(\gamma\mid\mathcal{H}, M) \propto \int p(\mathcal{H}\mid\gamma, \theta, M)  p(\gamma, \theta \mid M)d\theta ,
\label{eq:px}
\end{equation}

\begin{equation}
p_g(\gamma\mid\mathcal{S}, M) \propto \int p(\mathcal{S}\mid\gamma, \phi, M) p(\gamma, \phi \mid M)d\phi,
\label{eq:pgw}
\end{equation}
where $p(\mathcal{H}|\gamma, \theta, M)$ and $p(\mathcal{S}|\gamma, \phi, M)$ are the likelihood function of X-ray and GW, respectively. $p(\gamma, \theta | M)$ and $p(\gamma, \phi | M)$ are the prior probability of X-ray and GW, respectively.

\section{Simulation and Results} \label{sec:result}

We simulate 100 samples to estimate the initial frequency and ellipticity of the SMNS and test the validity of our method. The X-ray and GW signal from an SMNS is simulated as the set of input values for all parameters ($\gamma, \theta, \phi$) and we fixed the distinct parameter of X-ray and GW, $\theta$=($\eta$=0.1, $\alpha$=5.0, $n$ = 5.0, $\Delta t$ = 200 s, $L_{em,0}$ = $2.6 \times 10^{48}$ erg $\cdot$ s$^{-1}$), $\phi$ = ($I = 3\times10^{38}$ kg$\cdot$ m$^2$, $\iota$=0), respectively \citep{MOLNVIK1985239, Zhang2006ApJ, Lv2017ApJ, Lv2020ApJL,2021ApJ...912...14Y}. We assume that samples are uniformly distributed within the space constrained by the detection horizon, such that $p(d) = 3d^2/d^3_{max}$ \citep{Moore2015CQGra}. In order for the SMNS to collapse into a BH within 1000s and for its generated GW to be detected by ET, we set $d_{max} = 10$ Mpc, the ellipticity to be log-uniformly distributed in [0.008, 0.01] \citep{Lasky2016MN, Aasi2014ApJ}, and the initial frequency to be uniform distribution in [500, 1000] Hz \citep{Lasky2015PASA, Corsi2011Prd}. The above parameters ensure that the total energy of GW and X-ray will not exceed the rotational energy, and also ensure that the GW radiation is dominant in the system.  In X-ray data simulation, the noise is Gaussian and stationary, $n_x \sim \mathcal{N}(\mu, \sigma_{EM}^2)$ and $\mu=1.0\times10^{-10}$,  $\sigma_{EM} = 0.01 \times \bar{F}$, $\bar{F}$ is the mean of $F(t)$. In GW data simulation, the instrument noise $n(t)$ is that of the ET detector. Because the strain of GW decayed with time, we select the first 20 seconds of data for analysis. The optimal signal-to-noise ratios (SNRs) of 100 samples are all within [8.53, 91.44].

 From the obtained X-ray and GW simulated data, we use Bayes' theorem and bilby (version 1.1.1, dynesty sampler) \citep{Romero2020MNRAS,Ashton2019ApJS} to parameter estimation of the initial frequency and ellipticity of the SMNS. Because of lack of the prior information on ellipticity and initial frequency, we set the prior probability of $\epsilon$ to be log-uniform in the range [$10^{-9}, 10^{-1}$] \citep{Bonazzola1996AA, Lasky2016MN, Aasi2014ApJ}, and the $f_0$ is uniform distribution in the range [100, 1000] Hz \citep{Lasky2016MN, Corsi2011Prd, Abbott2021Prd}.

\begin{figure}
    \centering
    \includegraphics[width=8cm]{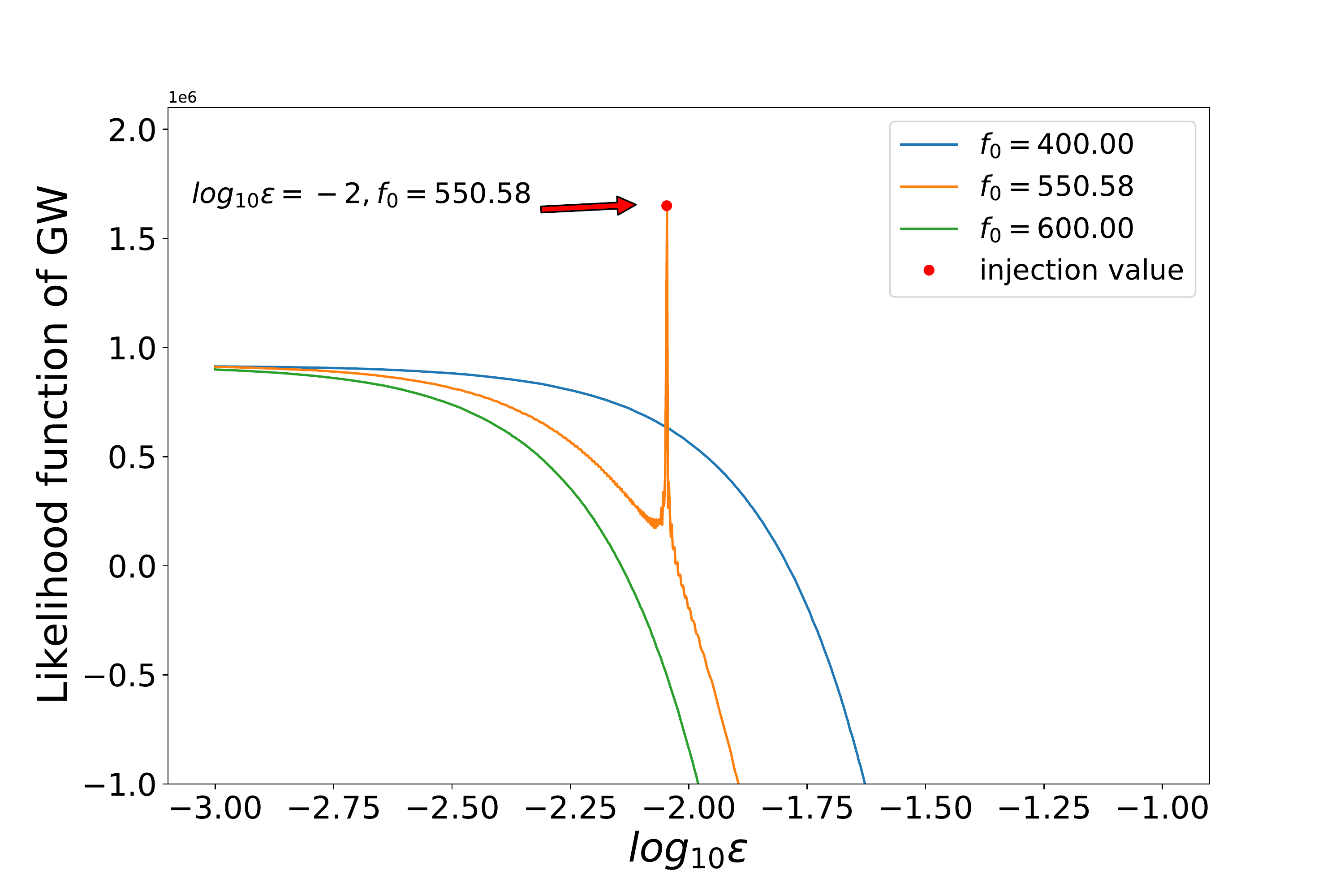}
    \caption{The likelihood function of GW under white Gaussian noise. The mean of noise is 0 and standard deviation is $8.38 \times 10^{-22}$. The test parameters for GW are $d=4.11$ Mpc, $I=3\times10^{38}$ kg$\cdot $m$^2$, $f_0=550.58$ Hz, $\epsilon=0.009$ and $\iota=0$.}
\label{fig:LK}
\end{figure}

According to Eq (\ref{eq:pgw}), when using the GW data alone to parameter estimate ellipticity and initial frequency, we find that the estimated  parameter distribution does not easily converge at the true value. We tested the likelihood function of GW under white Gaussian noise, and the result is shown in Figure \ref{fig:LK}.  The plausible reason for this parameter estimation results is that we lack the accurate prior information about ellipticity and initial frequency \footnote{The parameter estimation in  continuous-wave study (e.g. see the review in \citep{Keith2022arXiv}) adopt much narrow prior than that in our paper.}, and the range of prior is relatively wide.

\begin{figure}
    \centering
       \includegraphics[width=8.5cm]{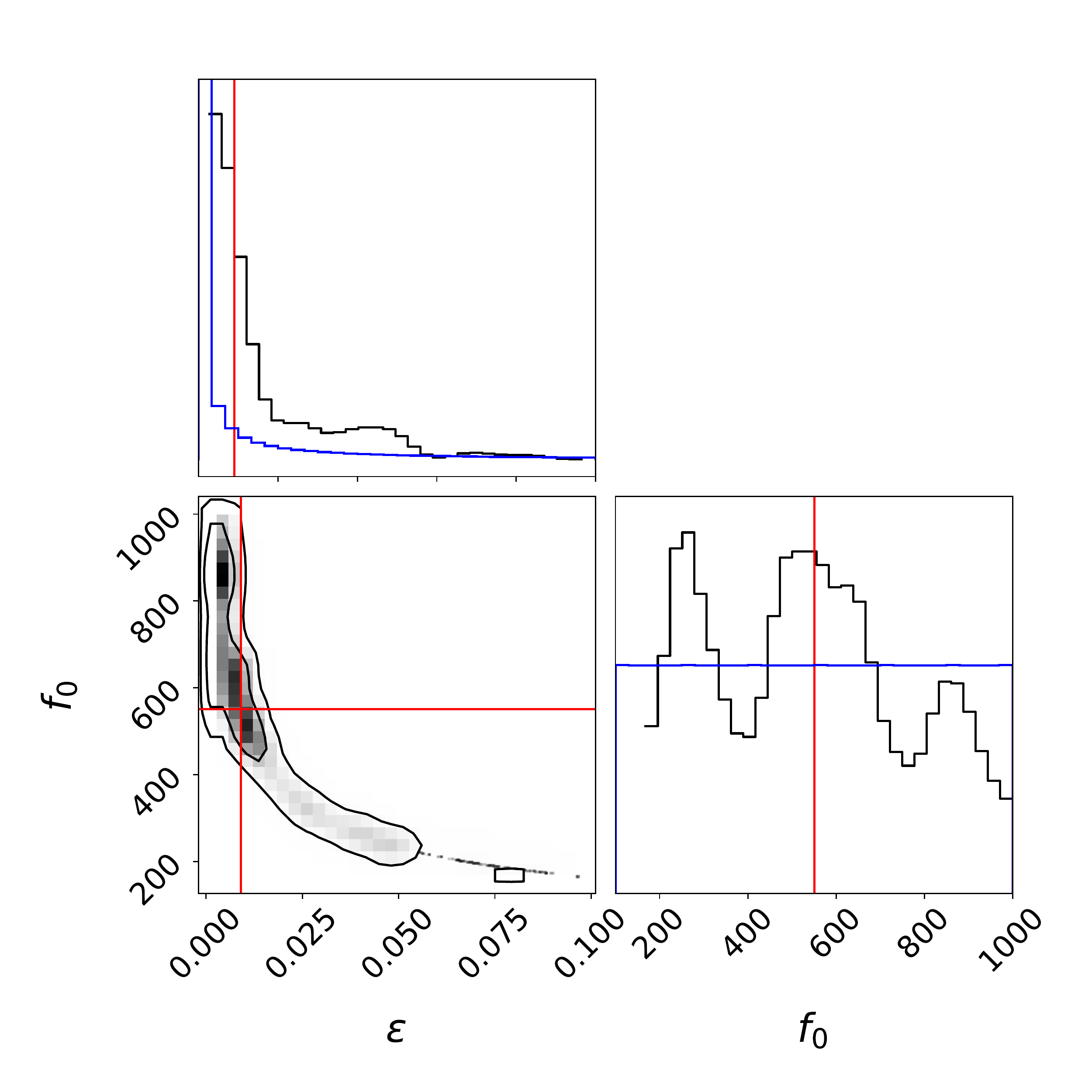}
    \includegraphics[width=8.5cm]{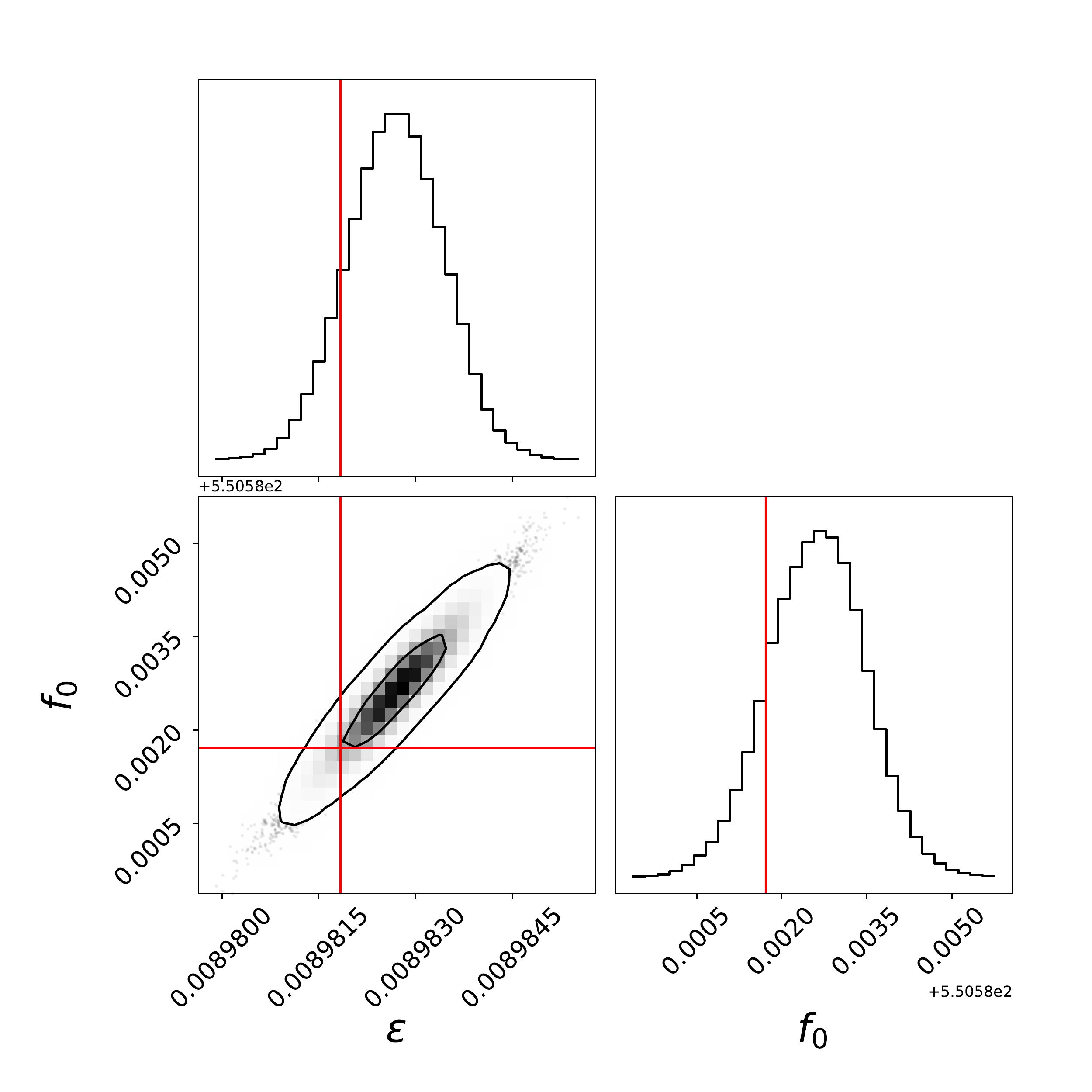}
    \caption{Posterior probability density on the initial frequency and ellipticity. The contours show $50\%$ and $95\%$ credible regions. We get the $\epsilon = 0.0094^{+0.0606}_{-0.0064}$ and $f_0 = 539.41^{+411.25}_{-341.89}$ Hz in $95\%$ credible regions using X-ray data alone (up-panle). While, $\epsilon = 0.0090^{+1\times10^{-6}}_{-2\times10^{-6}}$ and $f_0 = 550.58^{+1.6\times10^{-3}}_{-1.7\times10^{-3}}$ Hz in $95\%$ credible regions with joint X-ray–GW analysis (low-panel). The blue lines represent prior distributions, the black lines represent the results of parameter estimation, and the red lines represent the true values.}
    \label{fig:d_e_f}
\end{figure}

Observing an X-ray event in conjunction with a GW provides additional constraints on the ellipticity and initial frequency which could reduce the uncertainty in the estimation of the parameters. Although the ellipticity and initial frequency are degenerate when using X-ray alone for parameter estimation, the range of the obtained posterior distribution is much smaller than the prior distribution (e.g. see the result of the case study shown in the up-panel of Figure \ref{fig:d_e_f}). Therefore,  we can use the posterior distribution estimated by X-ray data as the prior for  GW data analysis, in order to solve the problem that parameter estimation with only GW data does not easily converge at the true value. In this approach,  Eq (\ref{eq:pgw}) can be rewritten as 
\begin{equation}
p_g(\gamma|\mathcal{S}, M) \propto \int p(\mathcal{S}|\gamma, \phi, M) p_x(\gamma|\mathcal{H}, M)  p(\phi | M) d\phi 
\label{eq:pgwx}
\end{equation}
where $p(\phi | M)$ is prior probability obtained by the X-ray data analysis. 

\begin{figure}
    \centering
    \includegraphics[width=7cm]{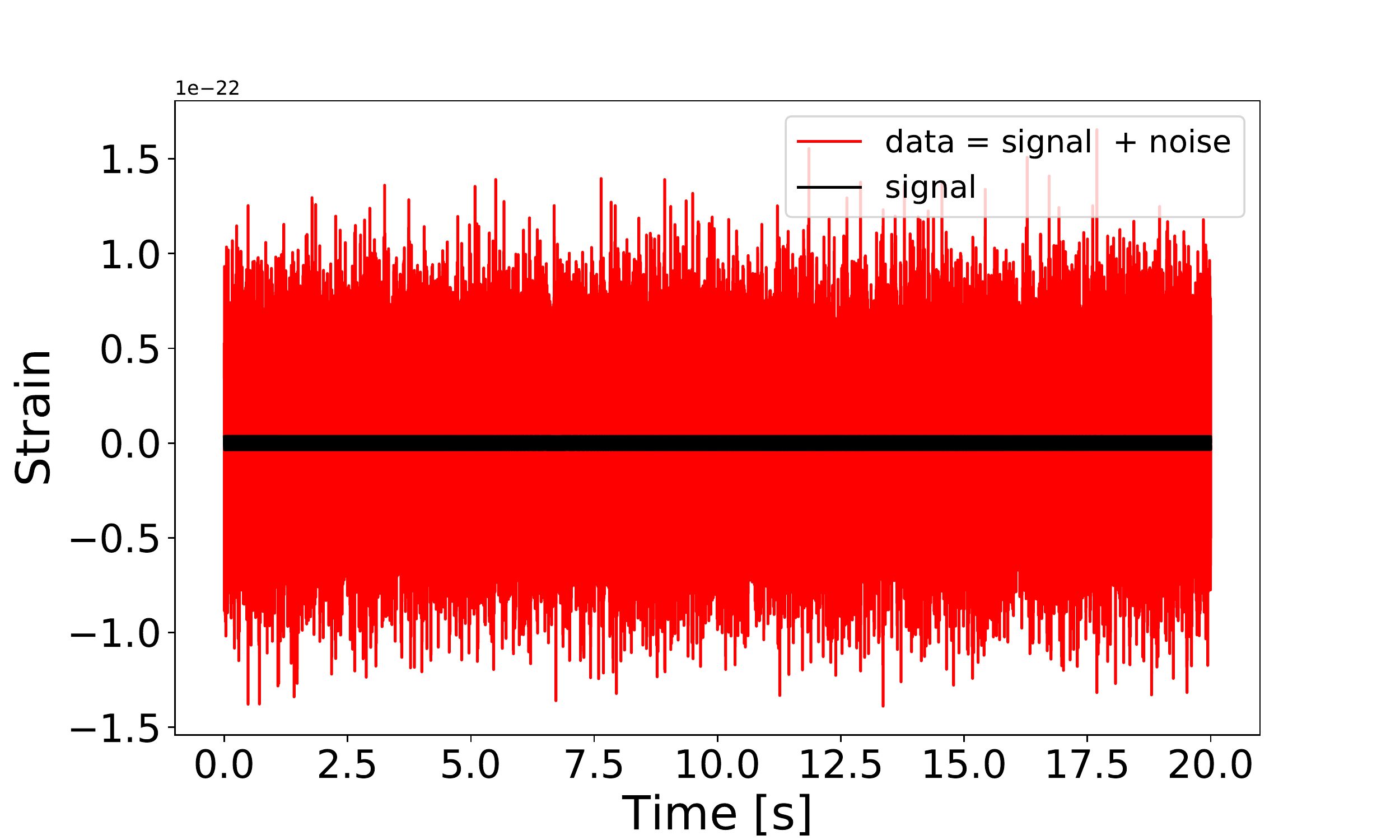}
    \includegraphics[width=7cm]{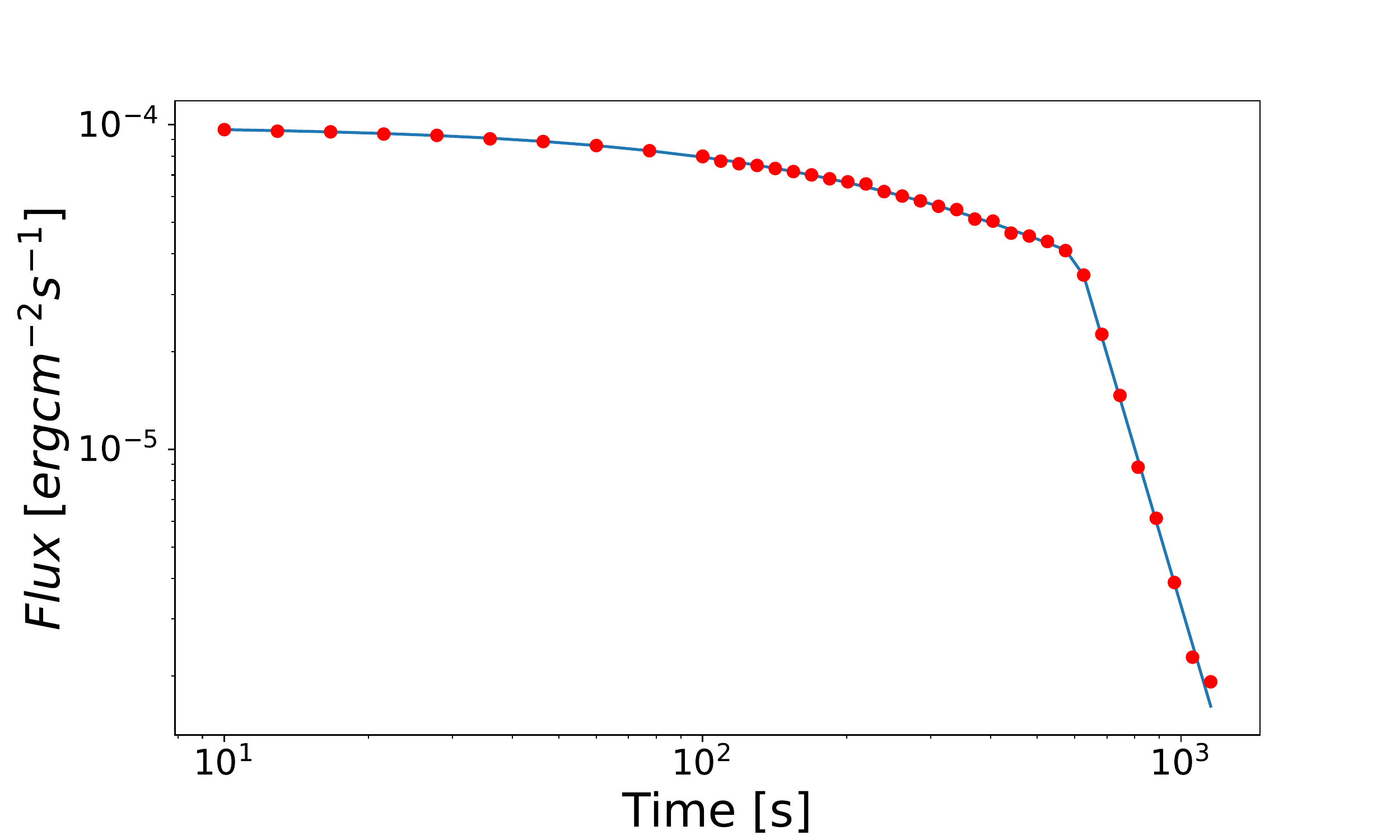}
    \caption{The simulated observation data at 4.11 Mpc, the initial frequency $f_0 = 550.58$ Hz and ellipticity $\epsilon = 0.0090$. The time 0s in the GW data corresponds to the trigger time of the X-ray detector. The upper panel shows the simulated GW data, with the red line representing the simulated observed GW data and the black line representing the simulated GW signal; and the lower panel shows the simulated EM data, with the red point representing the simulated observed data and the black line representing the model.}
    \label{fig:data}
\end{figure}

We select one of the simulations as a case study for illustrating the joint analysis. The parameter value of X-ray and GW for this sample is shown in Table \ref{tab:value}, and the value of SNR is 20.08 for this case. The time-domain signals of GW and X-ray are shown in Figure \ref{fig:data}. The time 0s in the GW data corresponds to the trigger time of the X-ray detector.

\begin{table*}
    \centering
    \caption{Parameter values used for X-ray and GW data simulations for the case study.}\label{tab:value}
    \begin{tabular}{cccccccccccc}
    \hline
	$\eta$ &	$\alpha$ & $n$ & $t_{col}$ (s) &{$L_{em,0}$ (erg $\cdot$ s$^{-1})$} &  {d (Mpc)} & {$f_0$ (Hz)} & {$\epsilon$} & {$I$ (kg$\cdot$ $m^2$)} & {$\iota$}\\
		\hline
	0.1 & 5.0 & 5.0 & 607.89 & $2.6\times10^{48}$ & 4.11 & 550.58 & 0.0090 & $3\times10^{38}$ & 0 \\
		\hline
    \end{tabular}
\end{table*}

The parameter estimation result by  the X-ray data alone  is shown in the up-panel of Figure \ref{fig:d_e_f}. The contours show $50\%$ and $95\%$ credible regions and the ellipticity and initial frequency are inferred to be $\epsilon = 0.0094^{+0.0606}_{-0.0064}$ and $f_0 = 539.41^{+411.25}_{-341.89}$ Hz in $95\%$ credible regions. This is predominantly a consequence of the degeneracy between the ellipticity and the initial frequency, and the result matches our expectations based on Eq (\ref{flux}). 

According to the parameter estimation of X-ray, we can obtain the posterior distributions of ellipticity and initial frequency, and their $95\%$ credible region is selected as the prior for the parameter estimation of GW data. The result of parameter estimation of GW data based on Eq (\ref{eq:pgwx}) is shown in the low-panel of Figure \ref{fig:d_e_f}, and the contour is shown the $50\%$ and $95\%$ credible regions. The ellipticity and initial frequency are inferred to be $\epsilon = 0.0090^{+1\times10^{-6}}_{-2\times10^{-6}}$ and $f_0 = 550.58^{+1.6\times10^{-3}}_{-1.7\times10^{-3}}$ Hz in $95\%$ credible regions. We find that the joint X-ray–GW  can not only reduce the range of parameters and greatly save the time of parameter estimation, but also effectively improve the accuracy of parameter estimation.

\begin{figure}
    \centering
    \includegraphics[width=8cm]{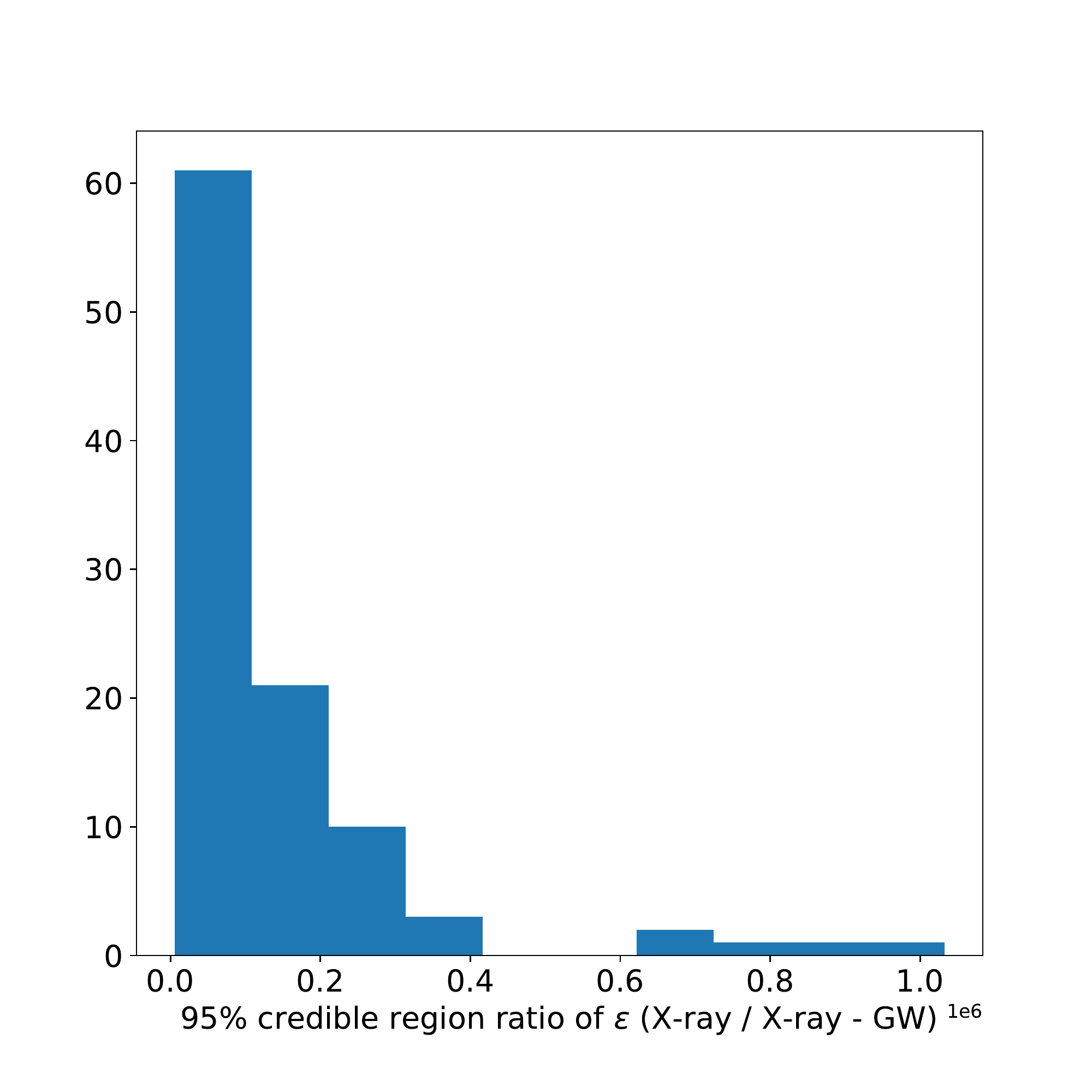}
    \includegraphics[width=8cm]{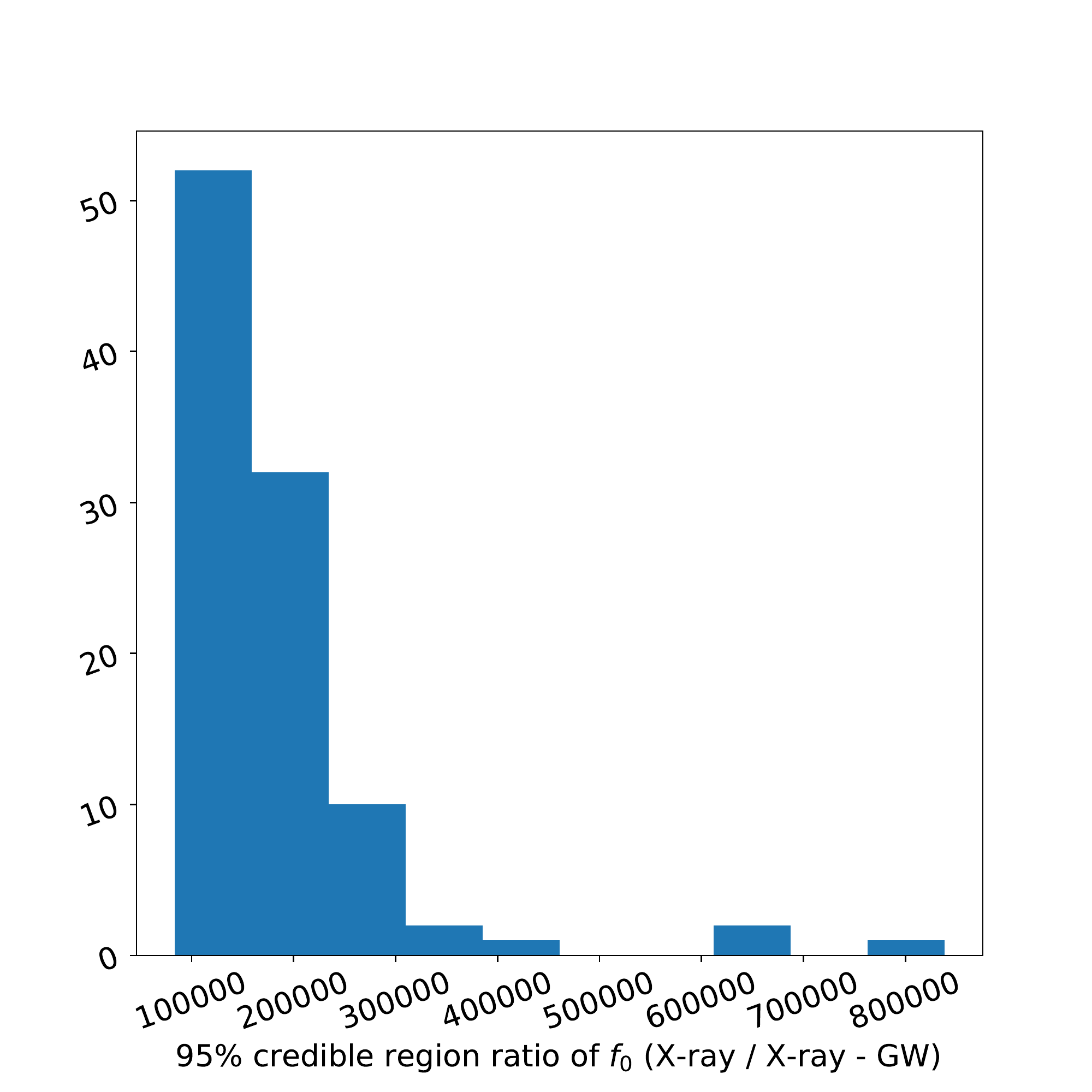}
    \caption{Histogram of the ratios of the 95$\%$ confidence intervals of the ellipiticity (up-panel) and initial frequency (low-panel) posterior distributions for X-ray and X-ray–GW for 100 samples.}
    \label{fig:hist_95}
\end{figure}

\begin{figure}
    \centering
    \includegraphics[width=8cm]{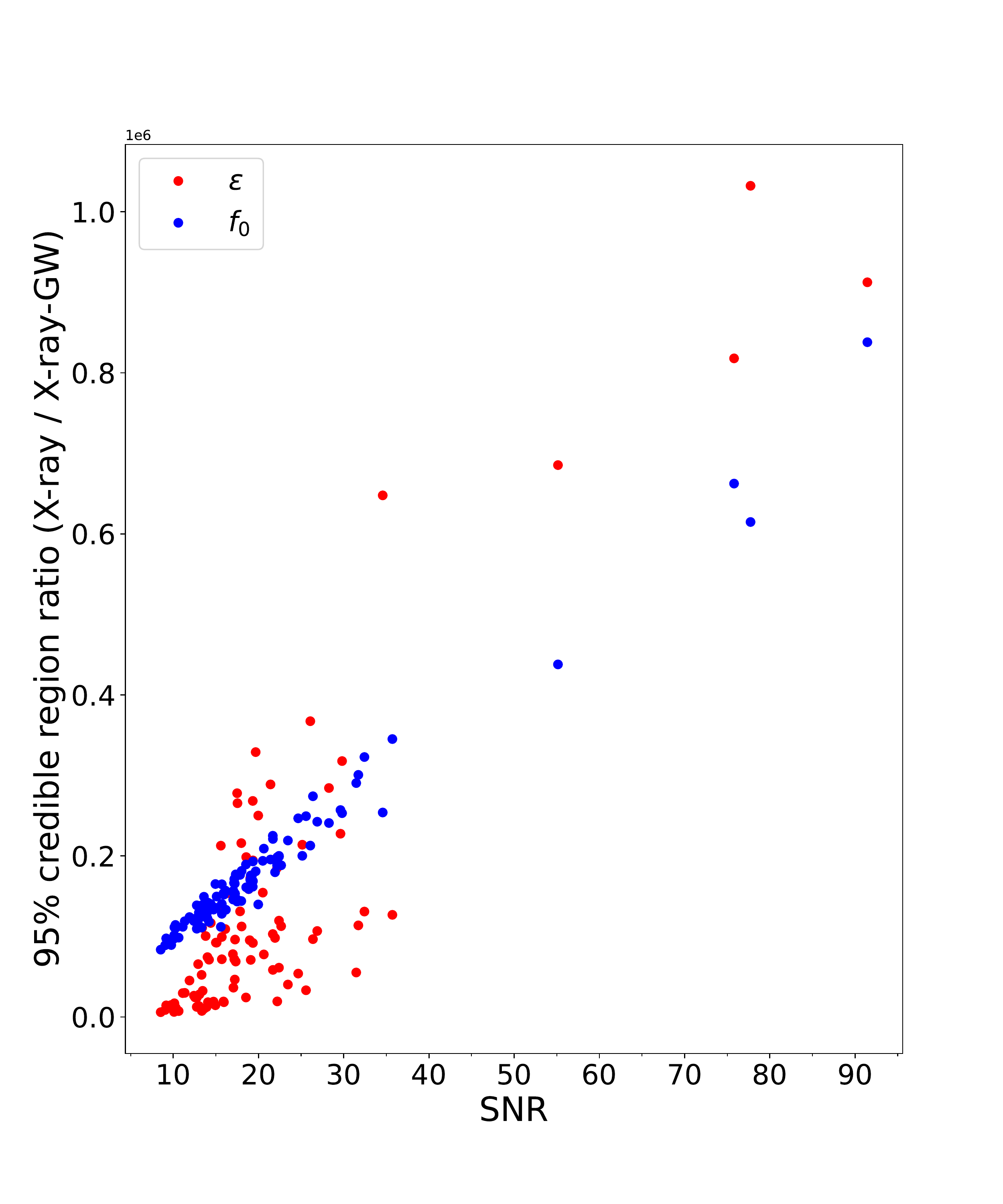}
    \caption{The correlation between $95\%$ confidence region ratio (X-ray / X-ray-GW) and SNR of GW signals  for the results of 100 samples shown in Fig \ref{fig:hist_95}.}
    \label{fig:snr_credible}
\end{figure}

We investigate the effectiveness of the joint X-ray–GW  analysis by examining the posterior credible intervals for the inferred initial frequency and ellipticity of the neutron star with 100 simulated signals as described previously. Shown in Figure \ref{fig:hist_95}, for joint  X-ray–GW analysis, the 95$\%$ credible regions for both ellipticity (up-panel) and initial frequency (low-panel) can be reduced to $\sim$ $10^{-5}$ of that for X-ray analysis alone. We investigated the correlation between the ratio of the $95\%$ confidence region (X-ray / X-ray-GW) and SNR for $\epsilon$ and $f_0$. The results are presented in Figure \ref{fig:snr_credible}, where it was observed that the ratio of the confidence region increases with the increase of SNR. Specifically, in the low SNR region (SNR $\leq$ 40), the $95\%$ confidence interval ratio of $f_0$ is greater than that of $\epsilon$, whereas in the high SNR region (SNR>40), the $95\%$ confidence interval ratio of $\epsilon$ is greater. Overall, the $95\%$ confidence interval ratio for $f_0$ exhibits a stronger correlation with SNR. In addition, we investigated the correlation between the $95\%$ confidence interval obtained by using X-ray data alone and the X-ray data error. We find that even when the error was artificially increased to 10 times of its original value, the ratio of the resulting $95\%$ confidence interval to the original $95\%$ confidence interval remained within a two-fold range.

\section{Summary} \label{sec:summary}
We simulate 100 samples of X-ray and GW data generated by the SMNS after the BNS merger and estimate the ellipticity and initial frequency of the SMNS by combining the X-ray and GW data. If the value of the ellipticity and initial frequency of the SMNS can exactly be estimated, then we could directly understand the inner structure of the SMNS. In addition to this, we have quantified the level to which the ellipticity-initial frequency degeneracy inherent to X-ray observations can be broken with the inclusion of GW data. According to the results shown in the up-panel of Figure \ref{fig:d_e_f} for the case study, we can find that it is impossible to obtain accurate ranges of ellipticity and initial frequency of the SMNS by using X-ray data alone for parameter estimation. However, the $95\%$ credible region of the parameter estimation results using X-ray data alone can be used as the prior information for parameter estimation of GW data, so as to improve the accuracy of the parameter estimation, and the result is shown in the low-panel of  Figure \ref{fig:d_e_f}. By combining the information from X-ray and GW channels through the multi-messenger Bayesian data analysis approach,  the $95\%$ credible region of parameter estimation for both ellipticity and initial frequency are all improved by a factor of $\sim10^5$ than X-ray data alone shown in Figure \ref{fig:hist_95}. As shown in Figure \ref{fig:snr_credible}, this factor increases with increasing SNR. In comparison to $95\%$ confidence interval  ratio of $\epsilon$, the $95\%$ confidence interval ratio of $f_0$ exhibits a stronger correlation with SNR.

In this paper, we consider the physical landscape of
SMNS that has evolved over a long period of time (e.g. more than 20s in our simulation). We simply assume that the ellipticity is a constant during this SMNS phase since the  evolution of ellipticity is not well modeled yet. We choose the ellipticity range of [0.008, 0.01]. The reason for this is that we consider the SMNS generated by the merger of binary neutron stars, which has a strong magnetic field (B $\sim 10^{17}$ G) so that it can reach this magnitude of ellipticity \citep{Suvorov2022PhRvD}.
In fact, the ellipticity may change over time due to the influence of magnetic fields, starquakes, and accretion, and when ellipticity changes with time, the evolution of frequency will also become more complex \citep{Giliberti2022MNRAS}. In addition, under different EoS of NS, the relationship between the mass and radius of NS is different, which will also affect the intensity of GW radiation \citep{2022HuangPrd}. All those complex progress are believe to have minor
effects on the scope of current work and needed to be investigate in future work. 

Our joint X-ray–GW data analysis  approach (e.g. Eq. 15) is also flexible to apply for  the phase of few tens of milliseconds just  after the merger, where the NS evolves even more dramatically. For this phase, people can only simulate the GW waveform and EM signals  through numerical relativity techniques (e.g. \cite{Radice2017ApJL}, \cite{2022Breschi}). During that phase, the ellipticity is likely to be larger, and the GW frequency could reach 2-4 kHz. Therefore ET might detect those sources up-to 100 Mpc \citep{Keith2022arXiv}. If we can combine electromagnetic waves with GW from the very early stages after the merger, it's a very important piece of the puzzle for understanding NS.



\section*{Acknowledgements}
We thank the
anonymous referees for valuable comments and suggestions that
helped us to improve the manuscript.
We thank Lin Lan, Yu-Feng Li, Zhi-Qiang You, and Shao-qi Hou for their discussion. This work is supported by the National Natural Science Foundation of China (grant No. 11922303,11922301) and the Fundamental Research Funds for the Central Universities (grant No. 2042022kf1182).

\section*{Data AVAILABILITY}

This theoretical study did not generate any new data.



\bibliography{mnras_template} 

\begin{thebibliography}{}
\makeatletter
\relax
\def\mn@urlcharsother{\let\do\@makeother \do\$\do\&\do\#\do\^\do\_\do\%\do\~}
\def\mn@doi{\begingroup\mn@urlcharsother \@ifnextchar [ {\mn@doi@}
  {\mn@doi@[]}}
\def\mn@doi@[#1]#2{\def\@tempa{#1}\ifx\@tempa\@empty \href
  {http://dx.doi.org/#2} {doi:#2}\else \href {http://dx.doi.org/#2} {#1}\fi
  \endgroup}
\def\mn@eprint#1#2{\mn@eprint@#1:#2::\@nil}
\def\mn@eprint@arXiv#1{\href {http://arxiv.org/abs/#1} {{\tt arXiv:#1}}}
\def\mn@eprint@dblp#1{\href {http://dblp.uni-trier.de/rec/bibtex/#1.xml}
  {dblp:#1}}
\def\mn@eprint@#1:#2:#3:#4\@nil{\def\@tempa {#1}\def\@tempb {#2}\def\@tempc
  {#3}\ifx \@tempc \@empty \let \@tempc \@tempb \let \@tempb \@tempa \fi \ifx
  \@tempb \@empty \def\@tempb {arXiv}\fi \@ifundefined
  {mn@eprint@\@tempb}{\@tempb:\@tempc}{\expandafter \expandafter \csname
  mn@eprint@\@tempb\endcsname \expandafter{\@tempc}}}

\bibitem[\protect\citeauthoryear{{Aasi} et~al.,}{{Aasi}
  et~al.}{2014}]{Aasi2014ApJ}
{Aasi} J.,  et~al., 2014, \mn@doi [\apj] {10.1088/0004-637X/785/2/119}, \href
  {https://ui.adsabs.harvard.edu/abs/2014ApJ...785..119A} {785, 119}

\bibitem[\protect\citeauthoryear{{Abbott} et~al.,}{{Abbott}
  et~al.}{2017a}]{2017CQGraAbbott_CE}
{Abbott} B.~P.,  et~al., 2017a, \mn@doi [Classical and Quantum Gravity]
  {10.1088/1361-6382/aa51f4}, \href
  {https://ui.adsabs.harvard.edu/abs/2017CQGra..34d4001A} {34, 044001}

\bibitem[\protect\citeauthoryear{Abbott et~al.,}{Abbott
  et~al.}{2017b}]{AbbottPRL}
Abbott B.~P.,  et~al., 2017b, \mn@doi [Phys. Rev. Lett.]
  {10.1103/PhysRevLett.119.161101}, 119, 161101

\bibitem[\protect\citeauthoryear{{Abbott} et~al.,}{{Abbott}
  et~al.}{2017c}]{2017ApJLAbbott}
{Abbott} B.~P.,  et~al., 2017c, \mn@doi [\apjl] {10.3847/2041-8213/aa9a35},
  \href {https://ui.adsabs.harvard.edu/abs/2017ApJ...851L..16A} {851, L16}

\bibitem[\protect\citeauthoryear{{Abbott} et~al.,}{{Abbott}
  et~al.}{2019}]{Abbott2019ApJ}
{Abbott} B.~P.,  et~al., 2019, \mn@doi [\apj] {10.3847/1538-4357/ab0f3d}, \href
  {https://ui.adsabs.harvard.edu/abs/2019ApJ...875..160A} {875, 160}

\bibitem[\protect\citeauthoryear{Abbott et~al.,}{Abbott
  et~al.}{2021}]{Abbott2021Prd}
Abbott R.,  et~al., 2021, \mn@doi [Phys. Rev. D] {10.1103/PhysRevD.104.082004},
  104, 082004

\bibitem[\protect\citeauthoryear{{Amati} et~al.,}{{Amati}
  et~al.}{2018}]{Amati2018}
{Amati} L.,  et~al., 2018, \mn@doi [Advances in Space Research]
  {10.1016/j.asr.2018.03.010}, \href
  {https://ui.adsabs.harvard.edu/abs/2018AdSpR..62..191A} {62, 191}

\bibitem[\protect\citeauthoryear{{Andersson}}{{Andersson}}{1998}]{1998ApJAndersson}
{Andersson} N.,  1998, \mn@doi [\apj] {10.1086/305919}, \href
  {https://ui.adsabs.harvard.edu/abs/1998ApJ...502..708A} {502, 708}

\bibitem[\protect\citeauthoryear{{Ashton} et~al.,}{{Ashton}
  et~al.}{2019}]{Ashton2019ApJS}
{Ashton} G.,  et~al., 2019, \mn@doi [\apjs] {10.3847/1538-4365/ab06fc}, \href
  {https://ui.adsabs.harvard.edu/abs/2019ApJS..241...27A} {241, 27}

\bibitem[\protect\citeauthoryear{{Baiotti}, {Giacomazzo}  \&
  {Rezzolla}}{{Baiotti} et~al.}{2008}]{Baiotti2008PhRvD}
{Baiotti} L.,  {Giacomazzo} B.,   {Rezzolla} L.,  2008, \mn@doi [\prd]
  {10.1103/PhysRevD.78.084033}, \href
  {https://ui.adsabs.harvard.edu/abs/2008PhRvD..78h4033B} {78, 084033}

\bibitem[\protect\citeauthoryear{{Barthelmy} et~al.,}{{Barthelmy}
  et~al.}{2005}]{Barthelmy2005Natur}
{Barthelmy} S.~D.,  et~al., 2005, \mn@doi [\nat] {10.1038/nature04392}, \href
  {https://ui.adsabs.harvard.edu/abs/2005Natur.438..994B} {438, 994}

\bibitem[\protect\citeauthoryear{{Beniamini}, {Duque}, {Daigne}  \&
  {Mochkovitch}}{{Beniamini} et~al.}{2020}]{Beniamini2020MNRAS}
{Beniamini} P.,  {Duque} R.,  {Daigne} F.,   {Mochkovitch} R.,  2020, \mn@doi
  [\mnras] {10.1093/mnras/staa070}, \href
  {https://ui.adsabs.harvard.edu/abs/2020MNRAS.492.2847B} {492, 2847}

\bibitem[\protect\citeauthoryear{{Bonazzola} \& {Gourgoulhon}}{{Bonazzola} \&
  {Gourgoulhon}}{1996}]{Bonazzola1996AA}
{Bonazzola} S.,  {Gourgoulhon} E.,  1996, \aap, \href
  {https://ui.adsabs.harvard.edu/abs/1996A&A...312..675B} {312, 675}

\bibitem[\protect\citeauthoryear{{Brady} \& {Creighton}}{{Brady} \&
  {Creighton}}{2000}]{Brady2000PhRvD}
{Brady} P.~R.,  {Creighton} T.,  2000, \mn@doi [\prd]
  {10.1103/PhysRevD.61.082001}, \href
  {https://ui.adsabs.harvard.edu/abs/2000PhRvD..61h2001B} {61, 082001}

\bibitem[\protect\citeauthoryear{{Breschi}, {Gamba}, {Borhanian}, {Carullo}  \&
  {Bernuzzi}}{{Breschi} et~al.}{2022}]{2022Breschi}
{Breschi} M.,  {Gamba} R.,  {Borhanian} S.,  {Carullo} G.,   {Bernuzzi} S.,
  2022, arXiv e-prints, \href
  {https://ui.adsabs.harvard.edu/abs/2022arXiv220509979B} {p. arXiv:2205.09979}

\bibitem[\protect\citeauthoryear{{Chatterjee} \& {Wen}}{{Chatterjee} \&
  {Wen}}{2022}]{Chatterjee2023}
{Chatterjee} C.,  {Wen} L.,  2022, \mn@doi [arXiv e-prints]
  {10.48550/arXiv.2301.03558}, \href
  {https://ui.adsabs.harvard.edu/abs/2023arXiv230103558C} {p. arXiv:2301.03558}

\bibitem[\protect\citeauthoryear{{Corsi} \& {M{\'e}sz{\'a}ros}}{{Corsi} \&
  {M{\'e}sz{\'a}ros}}{2009}]{Corsi2009ApJ}
{Corsi} A.,  {M{\'e}sz{\'a}ros} P.,  2009, \mn@doi [apj]
  {10.1088/0004-637X/702/2/1171}, \href
  {https://ui.adsabs.harvard.edu/abs/2009ApJ...702.1171C} {702, 1171}

\bibitem[\protect\citeauthoryear{{Corsi} \& {Owen}}{{Corsi} \&
  {Owen}}{2011}]{Corsi2011Prd}
{Corsi} A.,  {Owen} B.~J.,  2011, \mn@doi [\prd] {10.1103/PhysRevD.83.104014},
  \href {https://ui.adsabs.harvard.edu/abs/2011PhRvD..83j4014C} {83, 104014}

\bibitem[\protect\citeauthoryear{{Coyne}, {Corsi}  \& {Owen}}{{Coyne}
  et~al.}{2016}]{Coyne2016Prd}
{Coyne} R.,  {Corsi} A.,   {Owen} B.~J.,  2016, \mn@doi [\prd]
  {10.1103/PhysRevD.93.104059}, \href
  {https://ui.adsabs.harvard.edu/abs/2016PhRvD..93j4059C} {93, 104059}

\bibitem[\protect\citeauthoryear{{Cutler} \& {Schutz}}{{Cutler} \&
  {Schutz}}{2005}]{Cutler2005PhRvD}
{Cutler} C.,  {Schutz} B.~F.,  2005, \mn@doi [\prd]
  {10.1103/PhysRevD.72.063006}, \href
  {https://ui.adsabs.harvard.edu/abs/2005PhRvD..72f3006C} {72, 063006}

\bibitem[\protect\citeauthoryear{{Dai} \& {Lu}}{{Dai} \&
  {Lu}}{1998a}]{Dai1998PhRvL}
{Dai} Z.~G.,  {Lu} T.,  1998a, \mn@doi [\prl] {10.1103/PhysRevLett.81.4301},
  \href {https://ui.adsabs.harvard.edu/abs/1998PhRvL..81.4301D} {81, 4301}

\bibitem[\protect\citeauthoryear{{Dai} \& {Lu}}{{Dai} \&
  {Lu}}{1998b}]{Dai1998a}
{Dai} Z.~G.,  {Lu} T.,  1998b, \aap, \href
  {https://ui.adsabs.harvard.edu/abs/1998A&A...333L..87D} {333, L87}

\bibitem[\protect\citeauthoryear{{Dai}, {Wang}, {Wu}  \& {Zhang}}{{Dai}
  et~al.}{2006}]{2006SciDai}
{Dai} Z.~G.,  {Wang} X.~Y.,  {Wu} X.~F.,   {Zhang} B.,  2006, \mn@doi [Science]
  {10.1126/science.1123606}, \href
  {https://ui.adsabs.harvard.edu/abs/2006Sci...311.1127D} {311, 1127}

\bibitem[\protect\citeauthoryear{{Dietrich}, {Coughlin}, {Pang}, {Bulla},
  {Heinzel}, {Issa}, {Tews}  \& {Antier}}{{Dietrich}
  et~al.}{2020}]{Dietrich2020Sci}
{Dietrich} T.,  {Coughlin} M.~W.,  {Pang} P. T.~H.,  {Bulla} M.,  {Heinzel} J.,
   {Issa} L.,  {Tews} I.,   {Antier} S.,  2020, \mn@doi [Science]
  {10.1126/science.abb4317}, \href
  {https://ui.adsabs.harvard.edu/abs/2020Sci...370.1450D} {370, 1450}

\bibitem[\protect\citeauthoryear{{Engel} et~al.,}{{Engel}
  et~al.}{2022}]{Engel2022arXiv}
{Engel} K.,  et~al., 2022, \mn@doi [arXiv e-prints]
  {10.48550/arXiv.2203.10074}, \href
  {https://ui.adsabs.harvard.edu/abs/2022arXiv220310074E} {p. arXiv:2203.10074}

\bibitem[\protect\citeauthoryear{{Fan} \& {Hendry}}{{Fan} \&
  {Hendry}}{2015}]{Fan2015arXiv}
{Fan} X.,  {Hendry} M.,  2015, \mn@doi [arXiv e-prints]
  {10.48550/arXiv.1509.06022}, \href
  {https://ui.adsabs.harvard.edu/abs/2015arXiv150906022F} {p. arXiv:1509.06022}

\bibitem[\protect\citeauthoryear{{Fan} \& {Xu}}{{Fan} \&
  {Xu}}{2006}]{2006MNRASFan}
{Fan} Y.-Z.,  {Xu} D.,  2006, \mn@doi [\mnras]
  {10.1111/j.1745-3933.2006.00217.x}, \href
  {https://ui.adsabs.harvard.edu/abs/2006MNRAS.372L..19F} {372, L19}

\bibitem[\protect\citeauthoryear{{Fan}, {Wu}  \& {Wei}}{{Fan}
  et~al.}{2013}]{Fan2013PhRvD}
{Fan} Y.-Z.,  {Wu} X.-F.,   {Wei} D.-M.,  2013, \mn@doi [\prd]
  {10.1103/PhysRevD.88.067304}, \href
  {https://ui.adsabs.harvard.edu/abs/2013PhRvD..88f7304F} {88, 067304}

\bibitem[\protect\citeauthoryear{{Fan}, {Messenger}  \& {Heng}}{{Fan}
  et~al.}{2014}]{Fan2014ApJ}
{Fan} X.,  {Messenger} C.,   {Heng} I.~S.,  2014, \mn@doi [\apj]
  {10.1088/0004-637X/795/1/43}, \href
  {https://ui.adsabs.harvard.edu/abs/2014ApJ...795...43F} {795, 43}

\bibitem[\protect\citeauthoryear{{Fan}, {Messenger}  \& {Heng}}{{Fan}
  et~al.}{2017}]{Fan2017PhRvL}
{Fan} X.,  {Messenger} C.,   {Heng} I.~S.,  2017, \mn@doi [\prl]
  {10.1103/PhysRevLett.119.181102}, \href
  {https://ui.adsabs.harvard.edu/abs/2017PhRvL.119r1102F} {119, 181102}

\bibitem[\protect\citeauthoryear{{Fern{\'a}ndez} \& {Metzger}}{{Fern{\'a}ndez}
  \& {Metzger}}{2016}]{2016ARNPSRodrigo}
{Fern{\'a}ndez} R.,  {Metzger} B.~D.,  2016, \mn@doi [Annual Review of Nuclear
  and Particle Science] {10.1146/annurev-nucl-102115-044819}, \href
  {https://ui.adsabs.harvard.edu/abs/2016ARNPS..66...23F} {66, 23}

\bibitem[\protect\citeauthoryear{Flanagan \& Hinderer}{Flanagan \&
  Hinderer}{2008}]{FlanaganPRD2008}
Flanagan E.~E.,  Hinderer T.,  2008, \mn@doi [Phys. Rev. D]
  {10.1103/PhysRevD.77.021502}, 77, 021502

\bibitem[\protect\citeauthoryear{{Gao} \& {Fan}}{{Gao} \&
  {Fan}}{2006}]{Gao2006ChJAA}
{Gao} W.-H.,  {Fan} Y.-Z.,  2006, \mn@doi [\cjaa] {10.1088/1009-9271/6/5/01},
  \href {https://ui.adsabs.harvard.edu/abs/2006ChJAA...6..513G} {6, 513}

\bibitem[\protect\citeauthoryear{{Gao}, {Zhang}  \& {L{\"u}}}{{Gao}
  et~al.}{2016}]{2016PhRvDGao}
{Gao} H.,  {Zhang} B.,   {L{\"u}} H.-J.,  2016, \mn@doi [\prd]
  {10.1103/PhysRevD.93.044065}, \href
  {https://ui.adsabs.harvard.edu/abs/2016PhRvD..93d4065G} {93, 044065}

\bibitem[\protect\citeauthoryear{{Giacomazzo} \& {Perna}}{{Giacomazzo} \&
  {Perna}}{2013}]{Giacomazzo2013ApJL}
{Giacomazzo} B.,  {Perna} R.,  2013, \mn@doi [\apjl]
  {10.1088/2041-8205/771/2/L26}, \href
  {https://ui.adsabs.harvard.edu/abs/2013ApJ...771L..26G} {771, L26}

\bibitem[\protect\citeauthoryear{{Giliberti} \& {Cambiotti}}{{Giliberti} \&
  {Cambiotti}}{2022}]{Giliberti2022MNRAS}
{Giliberti} E.,  {Cambiotti} G.,  2022, \mn@doi [\mnras]
  {10.1093/mnras/stac245}, \href
  {https://ui.adsabs.harvard.edu/abs/2022MNRAS.511.3365G} {511, 3365}

\bibitem[\protect\citeauthoryear{{Goldstein} et~al.,}{{Goldstein}
  et~al.}{2017}]{Goldstein2017ApJL}
{Goldstein} A.,  et~al., 2017, \mn@doi [\apjl] {10.3847/2041-8213/aa8f41},
  \href {https://ui.adsabs.harvard.edu/abs/2017ApJ...848L..14G} {848, L14}

\bibitem[\protect\citeauthoryear{{Guidorzi} et~al.,}{{Guidorzi}
  et~al.}{2017}]{Guidorzi2017ApJ}
{Guidorzi} C.,  et~al., 2017, \mn@doi [\apjl] {10.3847/2041-8213/aaa009}, \href
  {https://ui.adsabs.harvard.edu/abs/2017ApJ...851L..36G} {851, L36}

\bibitem[\protect\citeauthoryear{{Haskell}, {Samuelsson}, {Glampedakis}  \&
  {Andersson}}{{Haskell} et~al.}{2008}]{Haskell2008MNRAS}
{Haskell} B.,  {Samuelsson} L.,  {Glampedakis} K.,   {Andersson} N.,  2008,
  \mn@doi [\mnras] {10.1111/j.1365-2966.2008.12861.x}, \href
  {https://ui.adsabs.harvard.edu/abs/2008MNRAS.385..531H} {385, 531}

\bibitem[\protect\citeauthoryear{{Hayes}, {Heng}, {Veitch}  \&
  {Williams}}{{Hayes} et~al.}{2020}]{Hayes2020ApJ}
{Hayes} F.,  {Heng} I.~S.,  {Veitch} J.,   {Williams} D.,  2020, \mn@doi [\apj]
  {10.3847/1538-4357/ab72fc}, \href
  {https://ui.adsabs.harvard.edu/abs/2020ApJ...891..124H} {891, 124}

\bibitem[\protect\citeauthoryear{{Hild}, {Chelkowski}  \& {Freise}}{{Hild}
  et~al.}{2008}]{2008Hild_ET}
{Hild} S.,  {Chelkowski} S.,   {Freise} A.,  2008, arXiv e-prints, \href
  {https://ui.adsabs.harvard.edu/abs/2008arXiv0810.0604H} {p. arXiv:0810.0604}

\bibitem[\protect\citeauthoryear{{Hotokezaka}, {Kiuchi}, {Kyutoku},
  {Muranushi}, {Sekiguchi}, {Shibata}  \& {Taniguchi}}{{Hotokezaka}
  et~al.}{2013}]{Hotokezaka2013PhRvD}
{Hotokezaka} K.,  {Kiuchi} K.,  {Kyutoku} K.,  {Muranushi} T.,  {Sekiguchi}
  Y.-i.,  {Shibata} M.,   {Taniguchi} K.,  2013, \mn@doi [\prd]
  {10.1103/PhysRevD.88.044026}, \href
  {https://ui.adsabs.harvard.edu/abs/2013PhRvD..88d4026H} {88, 044026}

\bibitem[\protect\citeauthoryear{{Huang}, {L{\"u}}, {Rice}  \& {Liang}}{{Huang}
  et~al.}{2022}]{2022HuangPrd}
{Huang} J.-X.,  {L{\"u}} H.-J.,  {Rice} J.,   {Liang} E.-W.,  2022, \mn@doi
  [\prd] {10.1103/PhysRevD.105.103019}, \href
  {https://ui.adsabs.harvard.edu/abs/2022PhRvD.105j3019H} {105, 103019}

\bibitem[\protect\citeauthoryear{{Kasen}, {Metzger}, {Barnes}, {Quataert}  \&
  {Ramirez-Ruiz}}{{Kasen} et~al.}{2017}]{Kasen2017Natur}
{Kasen} D.,  {Metzger} B.,  {Barnes} J.,  {Quataert} E.,   {Ramirez-Ruiz} E.,
  2017, \mn@doi [\nat] {10.1038/nature24453}, \href
  {https://ui.adsabs.harvard.edu/abs/2017Natur.551...80K} {551, 80}

\bibitem[\protect\citeauthoryear{{Kawaguchi}, {Shibata}  \&
  {Tanaka}}{{Kawaguchi} et~al.}{2018}]{Kawaguchi2018ApJL}
{Kawaguchi} K.,  {Shibata} M.,   {Tanaka} M.,  2018, \mn@doi [\apjl]
  {10.3847/2041-8213/aade02}, \href
  {https://ui.adsabs.harvard.edu/abs/2018ApJ...865L..21K} {865, L21}

\bibitem[\protect\citeauthoryear{{Lasky}}{{Lasky}}{2015}]{Lasky2015PASA}
{Lasky} P.~D.,  2015, \mn@doi [\pasa] {10.1017/pasa.2015.35}, \href
  {https://ui.adsabs.harvard.edu/abs/2015PASA...32...34L} {32, e034}

\bibitem[\protect\citeauthoryear{{Lasky} \& {Glampedakis}}{{Lasky} \&
  {Glampedakis}}{2016}]{Lasky2016MN}
{Lasky} P.~D.,  {Glampedakis} K.,  2016, \mn@doi [\mnras]
  {10.1093/mnras/stw435}, \href
  {https://ui.adsabs.harvard.edu/abs/2016MNRAS.458.1660L} {458, 1660}

\bibitem[\protect\citeauthoryear{{Lasky}, {Leris}, {Rowlinson}  \&
  {Glampedakis}}{{Lasky} et~al.}{2017}]{Lasky2017ApJ}
{Lasky} P.~D.,  {Leris} C.,  {Rowlinson} A.,   {Glampedakis} K.,  2017, \mn@doi
  [\apjl] {10.3847/2041-8213/aa79a7}, \href
  {https://ui.adsabs.harvard.edu/abs/2017ApJ...843L...1L} {843, L1}

\bibitem[\protect\citeauthoryear{{Lei}, {Wang}, {Zhang}, {Gan}, {Zou}  \&
  {Xie}}{{Lei} et~al.}{2009}]{2009ApJLei}
{Lei} W.~H.,  {Wang} D.~X.,  {Zhang} L.,  {Gan} Z.~M.,  {Zou} Y.~C.,   {Xie}
  Y.,  2009, \mn@doi [\apj] {10.1088/0004-637X/700/2/1970}, \href
  {https://ui.adsabs.harvard.edu/abs/2009ApJ...700.1970L} {700, 1970}

\bibitem[\protect\citeauthoryear{{Lindblom}, {Owen}  \& {Morsink}}{{Lindblom}
  et~al.}{1998}]{1998PhRvLLindblom}
{Lindblom} L.,  {Owen} B.~J.,   {Morsink} S.~M.,  1998, \mn@doi [\prl]
  {10.1103/PhysRevLett.80.4843}, \href
  {https://ui.adsabs.harvard.edu/abs/1998PhRvL..80.4843L} {80, 4843}

\bibitem[\protect\citeauthoryear{{Liu}, {Gu}  \& {Zhang}}{{Liu}
  et~al.}{2017}]{2017NewARLiu}
{Liu} T.,  {Gu} W.-M.,   {Zhang} B.,  2017, \mn@doi [\nar]
  {10.1016/j.newar.2017.07.001}, \href
  {https://ui.adsabs.harvard.edu/abs/2017NewAR..79....1L} {79, 1}

\bibitem[\protect\citeauthoryear{{L{\"u}}, {Zhang}, {Lei}, {Li}  \&
  {Lasky}}{{L{\"u}} et~al.}{2015}]{Lv2015ApJ}
{L{\"u}} H.-J.,  {Zhang} B.,  {Lei} W.-H.,  {Li} Y.,   {Lasky} P.~D.,  2015,
  \mn@doi [\apj] {10.1088/0004-637X/805/2/89}, \href
  {https://ui.adsabs.harvard.edu/abs/2015ApJ...805...89L} {805, 89}

\bibitem[\protect\citeauthoryear{{L{\"u}}, {Zhang}, {Zhong}, {Hou}, {Sun},
  {Rice}  \& {Liang}}{{L{\"u}} et~al.}{2017}]{Lv2017ApJ}
{L{\"u}} H.-J.,  {Zhang} H.-M.,  {Zhong} S.-Q.,  {Hou} S.-J.,  {Sun} H.,
  {Rice} J.,   {Liang} E.-W.,  2017, \mn@doi [\apj]
  {10.3847/1538-4357/835/2/181}, \href
  {https://ui.adsabs.harvard.edu/abs/2017ApJ...835..181L} {835, 181}

\bibitem[\protect\citeauthoryear{{L{\"u}}, {Zou}, {Lan}  \& {Liang}}{{L{\"u}}
  et~al.}{2018}]{2018MNRASLv}
{L{\"u}} H.-J.,  {Zou} L.,  {Lan} L.,   {Liang} E.-W.,  2018, \mn@doi [\mnras]
  {10.1093/mnras/sty2176}, \href
  {https://ui.adsabs.harvard.edu/abs/2018MNRAS.480.4402L} {480, 4402}

\bibitem[\protect\citeauthoryear{{L{\"u}}, {Lan}  \& {Liang}}{{L{\"u}}
  et~al.}{2019}]{Lv2019ApJ}
{L{\"u}} H.-J.,  {Lan} L.,   {Liang} E.-W.,  2019, \mn@doi [\apj]
  {10.3847/1538-4357/aaf71d}, \href
  {https://ui.adsabs.harvard.edu/abs/2019ApJ...871...54L} {871, 54}

\bibitem[\protect\citeauthoryear{{L{\"u}} et~al.,}{{L{\"u}}
  et~al.}{2020}]{Lv2020ApJL}
{L{\"u}} H.-J.,  et~al., 2020, \mn@doi [\apjl] {10.3847/2041-8213/aba1ed},
  \href {https://ui.adsabs.harvard.edu/abs/2020ApJ...898L...6L} {898, L6}

\bibitem[\protect\citeauthoryear{{Margalit} \& {Metzger}}{{Margalit} \&
  {Metzger}}{2017}]{Margalit2017ApJL}
{Margalit} B.,  {Metzger} B.~D.,  2017, \mn@doi [\apjl]
  {10.3847/2041-8213/aa991c}, \href
  {https://ui.adsabs.harvard.edu/abs/2017ApJ...850L..19M} {850, L19}

\bibitem[\protect\citeauthoryear{{M{\'e}sz{\'a}ros}, {Fox}, {Hanna}  \&
  {Murase}}{{M{\'e}sz{\'a}ros} et~al.}{2019}]{Meszaros2019NatRP}
{M{\'e}sz{\'a}ros} P.,  {Fox} D.~B.,  {Hanna} C.,   {Murase} K.,  2019, \mn@doi
  [Nature Reviews Physics] {10.1038/s42254-019-0101-z}, \href
  {https://ui.adsabs.harvard.edu/abs/2019NatRP...1..585M} {1, 585}

\bibitem[\protect\citeauthoryear{{Metzger}}{{Metzger}}{2017}]{Metzger2017LRR}
{Metzger} B.~D.,  2017, \mn@doi [Living Reviews in Relativity]
  {10.1007/s41114-017-0006-z}, \href
  {https://ui.adsabs.harvard.edu/abs/2017LRR....20....3M} {20, 3}

\bibitem[\protect\citeauthoryear{{Metzger} \& {Piro}}{{Metzger} \&
  {Piro}}{2014}]{Metzger2014MNRAS}
{Metzger} B.~D.,  {Piro} A.~L.,  2014, \mn@doi [\mnras] {10.1093/mnras/stu247},
  \href {https://ui.adsabs.harvard.edu/abs/2014MNRAS.439.3916M} {439, 3916}

\bibitem[\protect\citeauthoryear{M{\o}lnvik \& {\O}stgaard}{M{\o}lnvik \&
  {\O}stgaard}{1985}]{MOLNVIK1985239}
M{\o}lnvik T.,  {\O}stgaard E.,  1985, \mn@doi [Nuclear Physics A]
  {https://doi.org/10.1016/0375-9474(85)90235-0}, 437, 239

\bibitem[\protect\citeauthoryear{{Moore}, {Cole}  \& {Berry}}{{Moore}
  et~al.}{2015}]{Moore2015CQGra}
{Moore} C.~J.,  {Cole} R.~H.,   {Berry} C.~P.~L.,  2015, \mn@doi [Classical and
  Quantum Gravity] {10.1088/0264-9381/32/1/015014}, \href
  {https://ui.adsabs.harvard.edu/abs/2015CQGra..32a5014M} {32, 015014}

\bibitem[\protect\citeauthoryear{{Oganesyan}, {Ascenzi}, {Branchesi},
  {Salafia}, {Dall'Osso}  \& {Ghirlanda}}{{Oganesyan}
  et~al.}{2020}]{Oganesyan2020ApJ}
{Oganesyan} G.,  {Ascenzi} S.,  {Branchesi} M.,  {Salafia} O.~S.,  {Dall'Osso}
  S.,   {Ghirlanda} G.,  2020, \mn@doi [\apj] {10.3847/1538-4357/ab8221}, \href
  {https://ui.adsabs.harvard.edu/abs/2020ApJ...893...88O} {893, 88}

\bibitem[\protect\citeauthoryear{{Pitkin}}{{Pitkin}}{2011}]{Pitkin2011MNRAS}
{Pitkin} M.,  2011, \mn@doi [\mnras] {10.1111/j.1365-2966.2011.18818.x}, \href
  {https://ui.adsabs.harvard.edu/abs/2011MNRAS.415.1849P} {415, 1849}

\bibitem[\protect\citeauthoryear{{Pletsch}}{{Pletsch}}{2010}]{Pletsch2010PhRvD}
{Pletsch} H.~J.,  2010, \mn@doi [\prd] {10.1103/PhysRevD.82.042002}, \href
  {https://ui.adsabs.harvard.edu/abs/2010PhRvD..82d2002P} {82, 042002}

\bibitem[\protect\citeauthoryear{{Pletsch}}{{Pletsch}}{2011}]{Pletsch2011PhRvD}
{Pletsch} H.~J.,  2011, \mn@doi [\prd] {10.1103/PhysRevD.83.122003}, \href
  {https://ui.adsabs.harvard.edu/abs/2011PhRvD..83l2003P} {83, 122003}

\bibitem[\protect\citeauthoryear{{Pletsch} \& {Allen}}{{Pletsch} \&
  {Allen}}{2009}]{Pletsch2009PhRvL}
{Pletsch} H.~J.,  {Allen} B.,  2009, \mn@doi [\prl]
  {10.1103/PhysRevLett.103.181102}, \href
  {https://ui.adsabs.harvard.edu/abs/2009PhRvL.103r1102P} {103, 181102}

\bibitem[\protect\citeauthoryear{{Popham}, {Woosley}  \& {Fryer}}{{Popham}
  et~al.}{1999}]{1999ApJPopham}
{Popham} R.,  {Woosley} S.~E.,   {Fryer} C.,  1999, \mn@doi [\apj]
  {10.1086/307259}, \href
  {https://ui.adsabs.harvard.edu/abs/1999ApJ...518..356P} {518, 356}

\bibitem[\protect\citeauthoryear{{Price} \& {Rosswog}}{{Price} \&
  {Rosswog}}{2006}]{2006SciPrice}
{Price} D.~J.,  {Rosswog} S.,  2006, \mn@doi [Science]
  {10.1126/science.1125201}, \href
  {https://ui.adsabs.harvard.edu/abs/2006Sci...312..719P} {312, 719}

\bibitem[\protect\citeauthoryear{{Radice}, {Bernuzzi}, {Del Pozzo}, {Roberts}
  \& {Ott}}{{Radice} et~al.}{2017}]{Radice2017ApJL}
{Radice} D.,  {Bernuzzi} S.,  {Del Pozzo} W.,  {Roberts} L.~F.,   {Ott} C.~D.,
  2017, \mn@doi [\apjl] {10.3847/2041-8213/aa775f}, \href
  {https://ui.adsabs.harvard.edu/abs/2017ApJ...842L..10R} {842, L10}

\bibitem[\protect\citeauthoryear{{Riles}}{{Riles}}{2022}]{Keith2022arXiv}
{Riles} K.,  2022, \mn@doi [arXiv e-prints] {10.48550/arXiv.2206.06447}, \href
  {https://ui.adsabs.harvard.edu/abs/2022arXiv220606447R} {p. arXiv:2206.06447}

\bibitem[\protect\citeauthoryear{{Romero-Shaw} et~al.,}{{Romero-Shaw}
  et~al.}{2020}]{Romero2020MNRAS}
{Romero-Shaw} I.~M.,  et~al., 2020, \mn@doi [\mnras] {10.1093/mnras/staa2850},
  \href {https://ui.adsabs.harvard.edu/abs/2020MNRAS.499.3295R} {499, 3295}

\bibitem[\protect\citeauthoryear{{Rosswog}, {Davies}, {Thielemann}  \&
  {Piran}}{{Rosswog} et~al.}{2000}]{2000A&ARosswog}
{Rosswog} S.,  {Davies} M.~B.,  {Thielemann} F.~K.,   {Piran} T.,  2000, \aap,
  \href {https://ui.adsabs.harvard.edu/abs/2000A&A...360..171R} {360, 171}

\bibitem[\protect\citeauthoryear{{Rowlinson} et~al.,}{{Rowlinson}
  et~al.}{2010}]{Rowlinson2010MNRAS}
{Rowlinson} A.,  et~al., 2010, \mn@doi [\mnras]
  {10.1111/j.1365-2966.2010.17354.x}, \href
  {https://ui.adsabs.harvard.edu/abs/2010MNRAS.409..531R} {409, 531}

\bibitem[\protect\citeauthoryear{{Rowlinson}, {O'Brien}, {Metzger}, {Tanvir}
  \& {Levan}}{{Rowlinson} et~al.}{2013}]{Rowlinson2013MNRAS}
{Rowlinson} A.,  {O'Brien} P.~T.,  {Metzger} B.~D.,  {Tanvir} N.~R.,   {Levan}
  A.~J.,  2013, \mn@doi [\mnras] {10.1093/mnras/sts683}, \href
  {https://ui.adsabs.harvard.edu/abs/2013MNRAS.430.1061R} {430, 1061}

\bibitem[\protect\citeauthoryear{{Ruffert}, {Janka}  \& {Schaefer}}{{Ruffert}
  et~al.}{1996}]{1996A&ARuffert}
{Ruffert} M.,  {Janka} H.~T.,   {Schaefer} G.,  1996, \aap, \href
  {https://ui.adsabs.harvard.edu/abs/1996A&A...311..532R} {311, 532}

\bibitem[\protect\citeauthoryear{{Sarin}, {Lasky}, {Sammut}  \&
  {Ashton}}{{Sarin} et~al.}{2018}]{Sarin2018PhRvD}
{Sarin} N.,  {Lasky} P.~D.,  {Sammut} L.,   {Ashton} G.,  2018, \mn@doi [\prd]
  {10.1103/PhysRevD.98.043011}, \href
  {https://ui.adsabs.harvard.edu/abs/2018PhRvD..98d3011S} {98, 043011}

\bibitem[\protect\citeauthoryear{{Sarin}, {Lasky}  \& {Ashton}}{{Sarin}
  et~al.}{2020a}]{Sarin2020Prd}
{Sarin} N.,  {Lasky} P.~D.,   {Ashton} G.,  2020a, \mn@doi [\prd]
  {10.1103/PhysRevD.101.063021}, \href
  {https://ui.adsabs.harvard.edu/abs/2020PhRvD.101f3021S} {101, 063021}

\bibitem[\protect\citeauthoryear{{Sarin}, {Lasky}  \& {Ashton}}{{Sarin}
  et~al.}{2020b}]{Sarin2020MNRAS}
{Sarin} N.,  {Lasky} P.~D.,   {Ashton} G.,  2020b, \mn@doi [\mnras]
  {10.1093/mnras/staa3090}, \href
  {https://ui.adsabs.harvard.edu/abs/2020MNRAS.499.5986S} {499, 5986}

\bibitem[\protect\citeauthoryear{{Savchenko} et~al.,}{{Savchenko}
  et~al.}{2017}]{Savchenko2017ApJL}
{Savchenko} V.,  et~al., 2017, \mn@doi [\apjl] {10.3847/2041-8213/aa8f94},
  \href {https://ui.adsabs.harvard.edu/abs/2017ApJ...848L..15S} {848, L15}

\bibitem[\protect\citeauthoryear{{Shapiro}}{{Shapiro}}{2000}]{Shapiro2000ApJ}
{Shapiro} S.~L.,  2000, \mn@doi [\apj] {10.1086/317209}, \href
  {https://ui.adsabs.harvard.edu/abs/2000ApJ...544..397S} {544, 397}

\bibitem[\protect\citeauthoryear{{Shapiro} \& {Teukolsky}}{{Shapiro} \&
  {Teukolsky}}{1983}]{1983JBAAShapiro}
{Shapiro} S.~L.,  {Teukolsky} S.~A.,  1983, Journal of the British Astronomical
  Association, \href {https://ui.adsabs.harvard.edu/abs/1983JBAA...93R.276S}
  {93, 276}

\bibitem[\protect\citeauthoryear{{Shibata} \& {Taniguchi}}{{Shibata} \&
  {Taniguchi}}{2006}]{Shibata2006PhRvD}
{Shibata} M.,  {Taniguchi} K.,  2006, \mn@doi [\prd]
  {10.1103/PhysRevD.73.064027}, \href
  {https://ui.adsabs.harvard.edu/abs/2006PhRvD..73f4027S} {73, 064027}

\bibitem[\protect\citeauthoryear{{Shibata} \& {Ury{\={u}}}}{{Shibata} \&
  {Ury{\={u}}}}{2000}]{2000PhRvDShibata}
{Shibata} M.,  {Ury{\={u}}} K.~{\={o}}.,  2000, \mn@doi [\prd]
  {10.1103/PhysRevD.61.064001}, \href
  {https://ui.adsabs.harvard.edu/abs/2000PhRvD..61f4001S} {61, 064001}

\bibitem[\protect\citeauthoryear{{Sowell}, {Corsi}  \& {Coyne}}{{Sowell}
  et~al.}{2019}]{Sowell2019Prd}
{Sowell} E.,  {Corsi} A.,   {Coyne} R.,  2019, \mn@doi [\prd]
  {10.1103/PhysRevD.100.124041}, \href
  {https://ui.adsabs.harvard.edu/abs/2019PhRvD.100l4041S} {100, 124041}

\bibitem[\protect\citeauthoryear{{Suvorov} \& {Glampedakis}}{{Suvorov} \&
  {Glampedakis}}{2022}]{Suvorov2022PhRvD}
{Suvorov} A.~G.,  {Glampedakis} K.,  2022, \mn@doi [\prd]
  {10.1103/PhysRevD.105.L061302}, \href
  {https://ui.adsabs.harvard.edu/abs/2022PhRvD.105f1302S} {105, L061302}

\bibitem[\protect\citeauthoryear{{Wheeler}, {Yi}, {H{\"o}flich}  \&
  {Wang}}{{Wheeler} et~al.}{2000}]{2000ApJWheeler}
{Wheeler} J.~C.,  {Yi} I.,  {H{\"o}flich} P.,   {Wang} L.,  2000, \mn@doi
  [\apj] {10.1086/309055}, \href
  {https://ui.adsabs.harvard.edu/abs/2000ApJ...537..810W} {537, 810}

\bibitem[\protect\citeauthoryear{{Xing}, {Centrella}  \& {McMillan}}{{Xing}
  et~al.}{1994}]{1994PhRvDXing}
{Xing} Z.,  {Centrella} J.~M.,   {McMillan} S. L.~W.,  1994, \mn@doi [\prd]
  {10.1103/PhysRevD.50.6247}, \href
  {https://ui.adsabs.harvard.edu/abs/1994PhRvD..50.6247X} {50, 6247}

\bibitem[\protect\citeauthoryear{{Yu}, {Zhang}  \& {Gao}}{{Yu}
  et~al.}{2013}]{2013ApJLYu}
{Yu} Y.-W.,  {Zhang} B.,   {Gao} H.,  2013, \mn@doi [\apjl]
  {10.1088/2041-8205/776/2/L40}, \href
  {https://ui.adsabs.harvard.edu/abs/2013ApJ...776L..40Y} {776, L40}

\bibitem[\protect\citeauthoryear{{Yuan}, {L{\"u}}, {Yuan}, {Ma}, {Lei}  \&
  {Liang}}{{Yuan} et~al.}{2021}]{2021ApJ...912...14Y}
{Yuan} Y.,  {L{\"u}} H.-J.,  {Yuan} H.-Y.,  {Ma} S.-B.,  {Lei} W.-H.,   {Liang}
  E.-W.,  2021, \mn@doi [\apj] {10.3847/1538-4357/abedb1}, \href
  {https://ui.adsabs.harvard.edu/abs/2021ApJ...912...14Y} {912, 14}

\bibitem[\protect\citeauthoryear{{Zhang}}{{Zhang}}{2013}]{Zhang2013ApJL}
{Zhang} B.,  2013, \mn@doi [\apjl] {10.1088/2041-8205/763/1/L22}, \href
  {https://ui.adsabs.harvard.edu/abs/2013ApJ...763L..22Z} {763, L22}

\bibitem[\protect\citeauthoryear{{Zhang} \& {M{\'e}sz{\'a}ros}}{{Zhang} \&
  {M{\'e}sz{\'a}ros}}{2001}]{Zhang2001ApJ}
{Zhang} B.,  {M{\'e}sz{\'a}ros} P.,  2001, \mn@doi [\apjl] {10.1086/320255},
  \href {https://ui.adsabs.harvard.edu/abs/2001ApJ...552L..35Z} {552, L35}

\bibitem[\protect\citeauthoryear{{Zhang}, {Fan}, {Dyks}, {Kobayashi},
  {M{\'e}sz{\'a}ros}, {Burrows}, {Nousek}  \& {Gehrels}}{{Zhang}
  et~al.}{2006}]{Zhang2006ApJ}
{Zhang} B.,  {Fan} Y.~Z.,  {Dyks} J.,  {Kobayashi} S.,  {M{\'e}sz{\'a}ros} P.,
  {Burrows} D.~N.,  {Nousek} J.~A.,   {Gehrels} N.,  2006, \mn@doi [\apj]
  {10.1086/500723}, \href
  {https://ui.adsabs.harvard.edu/abs/2006ApJ...642..354Z} {642, 354}

\bibitem[\protect\citeauthoryear{{Zimmermann} \& {Szedenits}}{{Zimmermann} \&
  {Szedenits}}{1979}]{1979PhRvDZimmermann}
{Zimmermann} M.,  {Szedenits} E. J.,  1979, \mn@doi [\prd]
  {10.1103/PhysRevD.20.351}, \href
  {https://ui.adsabs.harvard.edu/abs/1979PhRvD..20..351Z} {20, 351}

\makeatother
\end{thebibliography}


\bsp	
\label{lastpage}
\end{document}